\documentclass[superscriptaddress,twocolumn]{revtex4}
\usepackage{amsmath,lscape,epsfig}
\usepackage{epstopdf,ulem}
\usepackage{color}

\def\beq{\begin{equation}}
\def\eeq{\end{equation}}
\def\beqa{\begin{eqnarray}}
\def\eeqa{\end{eqnarray}} 
\def\ban{\begin{eqnarray*}}
\def\ean{\end{eqnarray*}}
\def\bi{\begin{itemize}}
\def\ei{\end{itemize}}

\def\f{\frac}

\begin{document}

\title{Neutrino diffusion in the pasta phase matter within the Thomas-Fermi approach}

\author{U. J. Furtado}
\email{ujfurtado@gmail.com}
\affiliation{Depto de F\'{\i}sica - CFM - Universidade Federal de Santa
Catarina  Florian\'opolis - SC - CP. 476 - CEP 88.040 - 900 - Brazil} 
\affiliation{CFisUC, Department of Physics, University of Coimbra, P3030-076 Coimbra, Portugal} 
\author{S. S. Avancini}
\email{sidney@fsc.ufsc.br}
\affiliation{Depto de F\'{\i}sica - CFM - Universidade Federal de Santa
Catarina  Florian\'opolis - SC - CP. 476 - CEP 88.040 - 900 - Brazil} 
\affiliation{CFisUC, Department of Physics, University of Coimbra, P3030-076 Coimbra, Portugal} 
\author{J. R. Marinelli}
\email{ricardo.marinelli@ufsc.br}
\affiliation{Depto de F\'{\i}sica - CFM - Universidade Federal de Santa
Catarina  Florian\'opolis - SC - CP. 476 - CEP 88.040 - 900 - Brazil} 
\author{W. Martarello}
\email{williams85@gmail.com}
\affiliation{Depto de F\'{\i}sica - CFM - Universidade Federal de Santa
Catarina  Florian\'opolis - SC - CP. 476 - CEP 88.040 - 900 - Brazil} 
\author{C. Provid$\hat{e}$ncia}
\email{cp@fis.uc.pt }
\affiliation{CFisUC, Department of Physics, University of Coimbra, P3030-076 Coimbra, Portugal}

\begin{abstract}

The behavior and properties of  neutrinos in non-uniform  nuclear matter, surrounded by electrons and other neutrinos are studied. 
The nuclear matter itself is modeled by the non-linear Walecka model, where the so-called nuclear pasta phase is described using 
the Thomas-Fermi approximation, solved in a Wigner-Seitz cell. We  obtain the total cross-section and mean-free path for the neutrinos, 
taking into account scattering and neutrino absorption, and compare the final results for two known kind of model parametrizations: 
one in which non-linear effects in the strong sector are explicitly written in the model Lagrangian and another one in which the 
coupling constants are density dependent. The solution for this problem is important for the understanding of neutrino diffusion 
in a newly born neutron star after a supernova explosion.

\end{abstract}

\maketitle

\vspace{0.50cm}
PACS number(s): {24.10.Jv,26.60.-c,25.30.Pt,21.65.Mn}
\vspace{0.50cm}

\section{Introduction}\label{intro}

Neutrinos are elementary particles that interact with other particles
only through the weak force, which makes its scattering by matter very unlikely. On the other hand, this pure weak force scattering can reveal some aspects of the structure of matter that 
other  stronger interactions can not.  Also, this feature represents an advantage in the sense that neutrinos carry information 
about the target along great distances, as is the case of neutrinos produced in the interior of stars that reach earth detectors.  
For those reasons, the study of neutrino interaction with matter is at
the same time very fruitful but very challenging.
In what refers to the scattering of neutrinos by hadronic matter, despite of the experimental as well as theoretical problems, 
some important progress has been made as in the case of the Karmen collaboration  \cite{carmen}  and Los Alamos results \cite{alamo} . 
More recently, long-baseline experiments are under way, like MiniBoone, NOvA and other similar accelerator experiments \cite{links}. 
Although the main purpose is to obtain further information on Standard Models and on neutrino oscillations,  a precise knowledge of the
neutrino-hadron interaction as well as the hadronic structure of the targets  \cite{german} is necessary.

Another important source of information comes from the physics of the supernova core collapse and the evolution of a neutron star. 
The important ingredient in this case is the propagation of neutrinos in such a medium, which contains not only baryons but other leptons. 
In particular, for baryonic densities below the nuclear saturation value, structures known as pasta phases  \cite{pasta} are expected, 
which in some sense resemble the baryonic structure in nuclei, but now
embedded in a neutron gas and an electronic distribution. 
  Actually, a 
delicate competition between Coulomb and surface energies determine the most favorable final inhomogeneous structure. The presence 
of such structures presumably have a non-negligible role in the neutrino diffusion which 
is a key ingredient for the modeling and simulation  of the core
collapse supernovae mechanisms \cite{horowitz2005,sonoda2007,alloy11}.

One way to obtain the pasta phase is to solve the
problem considering charge neutral  Wigner-Seitz cells of appropriate geometries containing neutrons, protons and electrons within a 
variational approach, using both relativistic and non-relativistic mean-field 
calculations. Most of the recent applications in this case  have used
the Thomas-Fermi approximation \cite{pasta,maru1,tomas},
but Hartree-Fock
calculations \cite{gogelein} and three dimensional Thomas-Fermi
calculations \cite{maru2} can also be found in the recent literature,
all within the  Wigner-Seitz approximation. Recently, it was shown in
\cite{okamoto2012}  that performing a pasta calculation taking a large enough cell to include several
 units of the pasta structures, different distributions of matter from the usual ones
 considered within the Wigner Seitz (WS)  approximation 
 could  be energetically favored in some density ranges.
Another approach to this problem that also goes beyond WS
approximation  is based in the 
so-called quantum molecular dynamics \cite{qmd1,qmd2,schneider}.

Here we follow the Thomas-Fermi results as described in \cite{sidney} in order to generate the pasta-phase structure, where a 
relativistic model lagrangian is the starting point. The self-consistent calculation is performed considering matter in 
$\beta$-equilibrium, where just protons, neutrons, electrons and neutrinos are present. The total neutrino cross-section for 
each kind of particle is calculated, as a function of the density, taken the corresponding  geometry  for the considered 
density, i.e., droplet, rod, slab{, tube} or bubble. The  neutrino mean-free path (NMFP) energy and temperature
dependence  for selected values of the density is also presented.

The  neutrino cross-section calculation includes the
  effects of strong interaction, accounted for in-medium mass and
  energy shifts and degeneracy effects, based on the formalism  previously developed for homogeneous nuclear matter
\cite{Praka}, with the difference that now the strong and
electromagnetic potentials as well as the nucleon masses and energies are position dependent.
Consequently, the result is similar to the free space transition
amplitude, and the uniform matter result becomes, straightforwardly, a particular
case of our expressions for the cross-section.

We will not discuss in the present paper the important
sources of suppression and enhancement due to in-medium correlation
\cite{prakash99}. Our main objective is to understand how
inhomogeneous matter affects the neutrino cross section through the
in-medium effects.

As in \cite{Praka},
we have considered elastic scattering and neutrino absorption in our derivation.  In other words, neutral current and charge 
changing processes are included and their relative importance to the
total cross-section is discussed.
In what follows, an outline of the formalism is presented in section \ref{formalism}, numerical 
results and discussion are provided in section \ref{results}, and the conclusions are in the final section \ref{conclusions}.
The details of the pasta phase calculation can be found in  
the cited references and some details of the cross-section calculation are shown in the Appendix.

\section{Formalism and Model parametrization}\label{formalism}

We start with a model Lagrangian density that includes electrons, neutrinos, nucleons, the sigma, omega, rho and delta meson 
fields and the electromagnetic interaction, given by \cite{tomas,clebson}:

\begin{equation}
\mathcal{L}=\sum_{i=p,n}\mathcal{L}_{i}\mathcal{\,+L}_{{\sigma }}\mathcal{+L}_{{\omega }}\mathcal{+L}_{{\rho }}\mathcal{+L}_{{\omega \rho }}%
\mathcal{+L}_{{\delta }}\mathcal{+L}_{{\gamma }}\mathcal{\,+L}_{{e }}\mathcal{\,+L}_{{\nu }},
\label{lagdelta}
\end{equation}
where the nucleon Lagrangian reads
\begin{equation}
\mathcal{L}_{i}=\bar{\psi}_{i}\left[ \gamma _{\mu }iD^{\mu }-M^{*}\right]
\psi _{i}  \label{lagnucl},
\end{equation}
with
\begin{eqnarray}
iD^{\mu } &=&i\partial ^{\mu }-\Gamma_{v}V^{\mu }-\frac{\Gamma_{\rho }}{2}{\boldsymbol{\tau}}%
\cdot \mathbf{b}^{\mu } - e \frac{1+\tau_3}{2}A^{\mu}, \label{Dmu} \\
M^{*} &=&M-\Gamma_{s}\phi-\Gamma_{\delta }{\boldsymbol{\tau}}\cdot \boldsymbol{\delta}.
\label{Mstar}
\end{eqnarray}

The meson and electromagnetic Lagrangian densities are
\begin{eqnarray*}
\mathcal{L}_{{\sigma }} &=&\frac{1}{2}\left( \partial _{\mu }\phi \partial %
^{\mu }\phi -m_{s}^{2}\phi ^{2}-\frac{\kappa}{3} \phi^3 -\frac{\lambda}{12} \phi^4 \right)  \\
\mathcal{L}_{{\omega }} &=&\frac{1}{2} \left(-\frac{1}{2} \Omega _{\mu \nu }
\Omega ^{\mu \nu }+ m_{v}^{2}V_{\mu }V^{\mu }+ \frac{\zeta}{12}g_v^4 (V_\mu V^\mu)^2 \right) \\
\mathcal{L}_{{\rho }} &=&\frac{1}{2} \left(-\frac{1}{2}
\mathbf{B}_{\mu \nu }\cdot \mathbf{B}^{\mu
\nu }+ m_{\rho }^{2}\mathbf{b}_{\mu }\cdot \mathbf{b}^{\mu } \right)\\
\mathcal{L}_{ {\delta }} &=&\frac{1}{2}(\partial _{\mu }\boldsymbol{\delta}%
\partial ^{\mu }\boldsymbol{\delta}-m_{\delta }^{2}{\boldsymbol{\delta}}^{2})\\
\mathcal{L}_{{\gamma }} &=&-\frac{1}{4}F _{\mu \nu }F^{\mu \nu }\\
\mathcal{L}_{{\omega \rho }} &=&\Lambda (g_\rho^2 \mathbf{b}_{\mu }\cdot \mathbf{b}^{\mu})(g_v^2(V_\mu V^\mu)) ,
\end{eqnarray*}


\noindent where $\Omega _{\mu \nu }=\partial _{\mu }V_{\nu }-\partial _{\nu }V_{\mu }$
, $\mathbf{B}_{\mu \nu }=\partial _{\mu }\mathbf{b}_{\nu }-\partial _{\nu }\mathbf{b}%
_{\mu }-\Gamma_{\rho }(\mathbf{b}_{\mu }\times \mathbf{b}_{\nu })$ and $F_{\mu \nu }=\partial 
_{\mu }A_{\nu }-\partial _{\nu }A_{\mu }$. The electromagnetic coupling constant is given by $e=\sqrt{4 \pi/137}$
 and $\boldsymbol {\tau}$ is the isospin operator. Finally, the electron and neutrino Lagrangian densities read:
\begin{eqnarray*} 
\mathcal{L}_{e}&=&\bar{\psi}_{e}\left[ \gamma _{\mu }(i\partial ^{\mu }+eA^{\mu})-m_e\right]
\psi _{e}\\
\mathcal{L}_{\nu}&=&\bar{\psi}_{\nu}\left[ i\gamma _{\mu }\partial ^{\mu }\right]
\psi _{\nu}. 
\end{eqnarray*}

 The weak boson contributions to the above Lagrangians, affect the solutions of the corresponding 
 Euler-Lagrange equations, in a completely negligible way, due to their huge masses and relatively small energy range, and
 are only relevant in order to obtain the neutrino cross sections, so they are not included in the above equations in the 
 determination of the equations of state.

 The solution of the corresponding Euler-Lagrange equations are obtained in the mean field self-consistent
 Thomas-Fermi (TF) approximation, as explained in references \cite{tomas,sidney}. We have considered two types of
 parametrizations: one in which the couplings $\Gamma_{i}$ are taken to be constants (NL) and another one in which they are
 density dependent (DD). In the last case, the terms proportional to $\kappa$, $\lambda$, $\zeta$ and $\Lambda$ are set equal to zero. Within TF the
 system is considered locally uniform and the main output are the densities, which are position dependent.
 Explicitly, we have for the baryonic, scalar, isoscalar, scalar-isoscalar, electron and neutrino densities:

\beq
\rho(\boldsymbol r)= \rho_p(\boldsymbol r) + \rho_n(\boldsymbol r)= \left\langle \hat{\psi}^{\dagger} \hat{\psi} \right\rangle; \nonumber
\eeq

\beq
\rho_s(\boldsymbol r)= \rho_{s_p}(\boldsymbol r) + \rho_{s_n}(\boldsymbol r)= \left\langle \hat{\bar{\psi}} \hat{\psi} \right\rangle; \nonumber
\eeq

\beq
\rho_3(\boldsymbol r)= \rho_p(\boldsymbol r) -\rho_n(\boldsymbol r) = \left\langle  \hat{\psi}^{\dagger} \tau_3 \hat{\psi} \right\rangle; \nonumber
\eeq

\beq
\rho_{s3}(\boldsymbol r)= \rho_{sp}(\boldsymbol r) -\rho_{sn}(\boldsymbol r) = \left\langle  \hat{\bar{\psi}} \tau_3 \hat{\psi} \right\rangle; \nonumber
\eeq

\beq
\rho_e(\boldsymbol r)= \left\langle  \hat{\psi}^{\dagger}_e  \hat{\psi}_e \right\rangle; \nonumber
\label{rhoe}
\eeq

\beq
\rho_\nu(\boldsymbol r)= \left\langle  \hat{\psi}^{\dagger}_\nu  \hat{\psi}_\nu \right\rangle; \nonumber
\label{rhonu}
\eeq

\noindent with
\beq \label{dens b}
\rho_i(\boldsymbol r) = \f{\gamma}{(2\pi)^3}\int d^3 k ~ (\eta_{ki} (T)-\bar \eta_{ki} (T)), \quad i=p,n,e,\nu;
\eeq

\beq \label{dens s}
\rho_{si}(\boldsymbol r) = \f{\gamma}{(2\pi)^3}\int d^3 k ~ \f{M_i^*}{E_i^*} (\eta_{ki} (T)+\bar \eta_{ki} (T)), \quad i=p,n;  
\eeq

\noindent where $E^*=\sqrt{k^2+M^{*2}}$, $\boldsymbol{k}$ is the momentum
and $\gamma=2$ is the spin multiplicity. For a given temperature $T$, the distributions are:
$$
\eta_{ki}(T)=\{\exp{[(E_i-\mu_i)/T]}+1\}^{-1};
$$
\beq
\bar \eta_{ki}(T)=\{\exp{[(\bar E_i-\bar\mu_i)/T]}+1\}^{-1};
\eeq
with $E_i~(\bar E_i)$ and $\mu_i~(\bar\mu_i)$ being the particle (antiparticle) energy and chemical potential, respectively.
The particle (antiparticle) energy depends on the mesonic fields and is position dependent.

Once the densities are determined we proceed for the calculation of the total neutrino cross-section ($\sigma$). We follow here 
the procedure discussed in \cite{Praka}. For a collision $1 + 2$ $\longrightarrow$ $3 + 4$ we may write:

\begin{align} \label{secao nucleon}
\sigma =& {G_F}^2\int d^3 r \int \f{d^3 k_2}{(2\pi)^3} \int \f{d^3 k_3}{(2\pi)^3} \int \f{d^3 k_4}{(2\pi)^3}\f{(2\pi)^4}{|\vec{v}_1 - \vec{v}_2|}
\cdot\nonumber\\
& \delta^4 (P_1+P_2-P_3-P_4)\eta_2(T) (1-\eta_3 (T)) (1-\eta_4 (T))  \cdot\nonumber\\
&\{({\cal V}+{\cal A})^2(1-v_2cos(\theta_{12}))(1-v_4cos(\theta_{34}))\nonumber\\
&+({\cal V}-{\cal A})^2(1-v_4cos(\theta_{14}))(1-v_2cos(\theta_{23}))\nonumber\\
&- \f{M^*_2 M^*_4}{E_2^* E_4^*}({\cal V}^2-{\cal A}^2)(1-cos(\theta_{13}))\}.
\end{align}

In the above expression $P_i$ is the particle four momentum,
\noindent $v_i=|\boldsymbol{k}_i|/E^*_i$, ${G_F}$ is the Fermi constant and ${\cal V}$, ${\cal A}$ are the weak current vector and 
 axial vector couplings,
respectively, and depend on the target particle and on the exchanged weak boson. For the neutrino-electron(neutrino) cross 
section we replace $M^{*}$ by $m_e$(zero).
The factors $(1-\eta_3 (T))$ and $(1-\eta_4 (T))$ are due to final state Pauli blocking effects.
We have considered here scattering through neutral current and charge-changing 
processes as well as neutrino absorption by the neutrons. The explicit expressions and definitions for each case are
 shown in the Appendix. Through the analysis of those expressions we may conclude that the integrand in the cross section 
 is $\boldsymbol{r}$ dependent, once the potentials and the effective mass are position dependent. Note that the same expression 
 can be used
 for the calculation of the cross section in the infinite nuclear matter, for which the potentials and effective mass are 
 not $\boldsymbol{r}$ dependent.
 
 The nucleon internal structure was taken into account in our calculation, just multiplying the
 couplings $\cal V$ and $\cal A$ by the vector and axial-vector nucleon form factors as explained,
 for instance, in \cite{INPC2013}. 

The neutrino mean-free-path is then obtained considering the
cross-section  for a Wigner-Seitz cell divided by its volume
\beq
\lambda = \left(\f{\sigma}{V}\right)^{-1}.
\eeq

\section{Results and Discussion}\label{results}

\begin{figure}
\begin{center}
\includegraphics[scale=0.70]{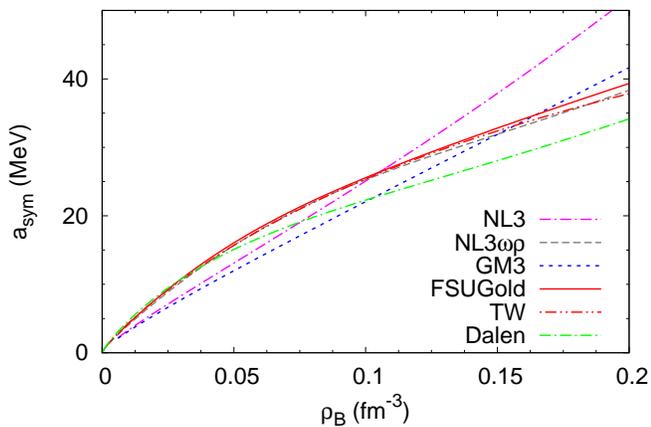}
\end{center}
\caption{Symmetry energy as a function of the total baryonic density
  for all the models used in the present work.}
\label{esym}
\end{figure}

In what follows we show our main results for the pasta phase, using both
Lagrangian parametrizations with constant and density dependent
couplings, as explained before.
For the parametrizations with non-linear terms in the mesonic sector
and constant couplings, 
we choose the GM3 \cite{gm3}, NL3 \cite{nl3}, NL3$\omega\rho$ \cite{hor01} and the FSUGold \cite{pika} sets
of parameters. The sets NL3 and NL3$\omega\rho$ only differ on the
density dependence of the symmetry energy and will allow the
discussion of the effect of this quantity on the neutrino mean-free path.
 For
the density dependent case we take the TW \cite{tw} and the van Dalen
{\it et al.} \cite{dalen} sets. The last one was recently 
taken for neutrino mean-free-path studies
\cite{gogelein}. Table \ref{tab para} shows the
  properties of the models for symmetric nuclear matter at saturation
  density $\rho_0$ and  zero
  temperature, and in Fig. \ref{esym} we have plotted the symmetry
  energy of the models versus the density, a quantity that will be
  necessary to discuss the results in the following. Models Dalen, GM3
  and NL3 have a smaller symmetry energy in all or part of the density
  range below 0.1 fm$^{-3}$. A smaller symmetry energy allows the
  system to have a larger isospin asymmetry. As we will see next these
  three models are precisely the ones with smaller proton fractions in
  the pasta phase matter, and this will induce a noticeable effect on
  the neutrino mean-free-path.

\begin{table}
  \centering
        \caption{Properties of the models parametrizations for infinite symmetric nuclear matter at zero temperature
at saturation density: the saturation density $\rho_0$, the binding energy $E_B/A$, the incompressibility $K$, the nucleon effective mass $M^\ast$, the symmetry energy $a_{sym}$ and its slope $L$.}
        \resizebox{8.7cm}{!}{
	\begin{tabular}{lcccccc}
                \hline
		 & $\rho_0$ & $E_B/A$ & $K$ & $M^\ast/M$ & $a_{sym}$ & $L$ \\
		 & (fm$^{-3})$ & (MeV) & (MeV) & & (MeV) & (MeV) \\
                \hline
		FSUGold & 0.148 & 16.299 & 271.76 & 0.6 & 37.4 & 60.4 \\
		GM3 & 0.153 & 16.3 & 240 & 0.78 & 32.5 & 89.66\\
		NL3 & 0.148 & 16.3 & 272 & 0.6 & 37.4 & 118.3 \\
		NL3$\omega\rho$ &  &  &  &  &  &  \\
                $(l_v=0.3)$ & 0.148 & 16.3 & 272 & 0.6 & 31.7 & 55.2 \\
		Dalen & 0.178 & 16.25 & 337 & 0.68 & 32.11 & 57 \\
		TW & 0.153 & 16.247 & 240 & 0.555 & 33.39 & 55.3 \\
		\hline
	\end{tabular}
  }
	\label{tab para}
\end{table}

\begin{figure}
\begin{center}
\includegraphics[scale=0.70]{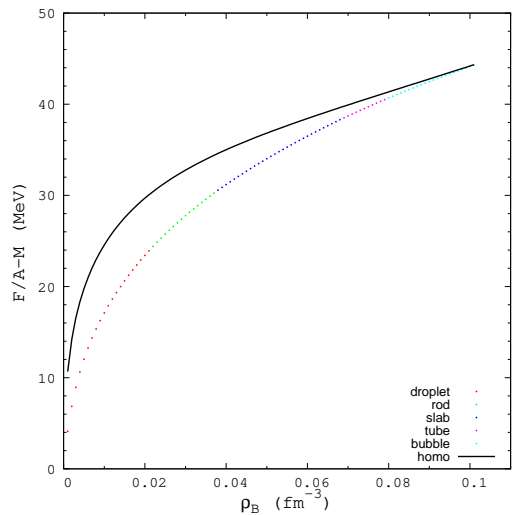}
\end{center}
\caption{Free energy as a function of the total baryonic density using the FSUGold parametrization for 
uniform matter (black) and the pasta phase matter within droplet
(red), rod (green), slab (blue), tube (purple) and bubble (light blue) geometries.
 The temperature was taken as $T=1$ MeV. $M$ is the free nucleon mass.}
\label{freeFST1}
\end{figure}

\begin{figure}
\begin{center}
\includegraphics[scale=0.70]{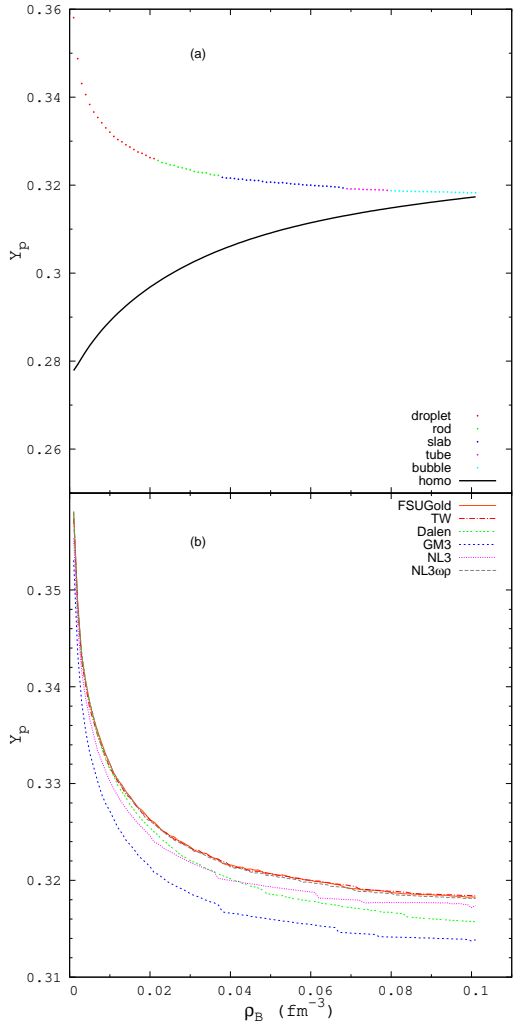}
\end{center}
\caption{Proton fraction as a function of the total baryonic density, (a) using the FSUGold parametrization for 
uniform matter (black) and the pasta phase matter within droplet
(red), rod (green), slab (blue), tube (purple) and bubble (light blue)
geometries, (b) for all the models and the pasta phase matter.
 The temperature was taken as $T=1$ MeV.}
\label{fracFST1}
\end{figure}

\begin{figure}
\begin{center}
\includegraphics[scale=0.70]{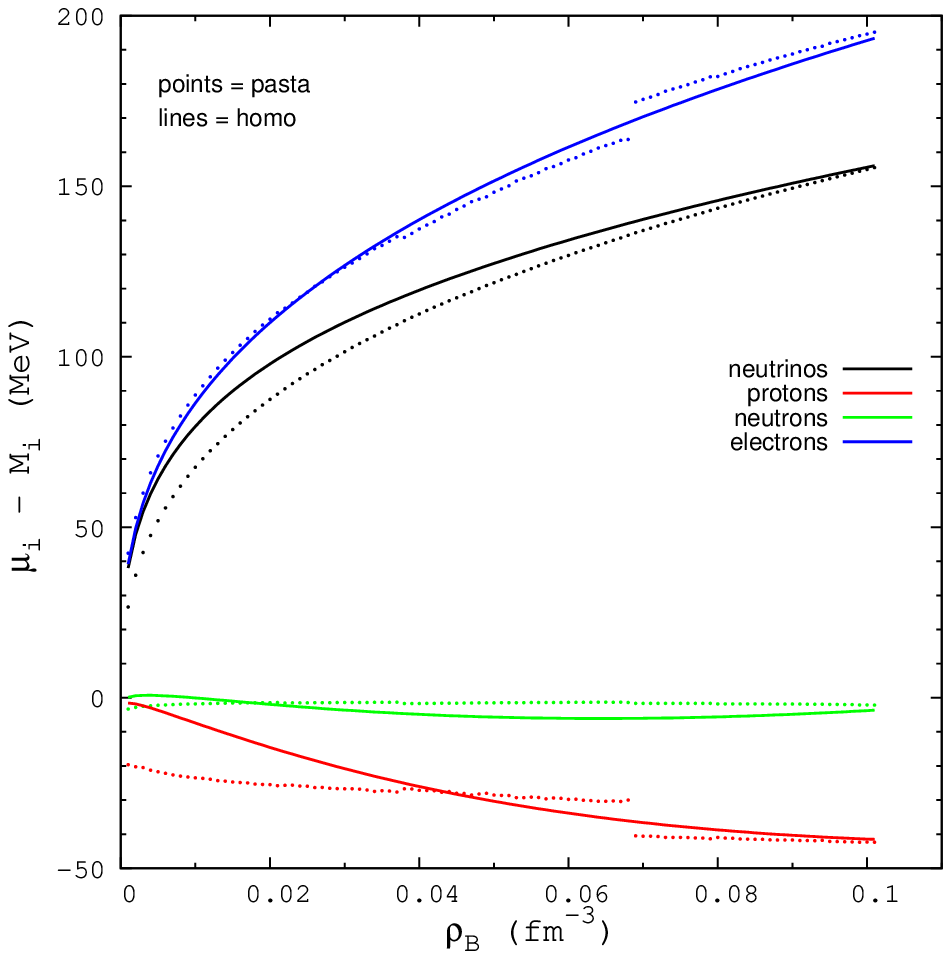}
\end{center}
\caption{Chemical potential as a function of the total baryonic density using the FSUGold parametrization for the
 pasta phase (points) compared to homogeneous matter (lines) for the
 neutrons (green), protons (red), electrons (blue) and neutrinos (black).
 The temperature was taken as $T=1$ MeV.}
\label{chemFST1}
\end{figure}

The free energy per particle obtained with the FSUGold
parametrization for $\beta$-equilibrium matter with trapped
  neutrinos for a fixed fraction of leptons $Y_L=0.4$ is shown in Fig. \ref{freeFST1}, where 
the different geometries are
identified by different colors. The uniform matter result is also
shown, and, as expected, has a larger free energy than the pasta phases. Although the transition densities between geometries may differ
slightly, depending on the parametrization used \cite{grill12}, the main behavior does not change significantly compared to the 
case shown.

Figures \ref{fracFST1}(a) and \ref{chemFST1} display, respectively, the proton fraction and the chemical potential for all particles 
involved using again the FSUGold parametrization to describe neutrino trapped
$\beta$-equilibrium matter. All these results were obtained for a
temperature $T=1$MeV. One important effect of the clusterization of
matter is the increase of the mean proton fraction.
An effect of the larger proton fraction
for matter with a fixed lepton fraction is the reduction of the
neutrino fraction and, consequently, the neutrino chemical
potential as seen in Fig. \ref{chemFST1}.  At low temperatures, when degeneracy effects are important,
 a smaller neutrino chemical potential will give rise to larger
mean-free paths, and neutrinos will diffuse out of the star more
easily. In Fig. \ref{fracFST1}(b) the pasta phase matter proton
fraction is plotted for all the models studied. These results reflect
the density dependence of the  symmetry energy of the respective
models: a smaller proton fraction occurs for the models with smaller
symmetry energy at the baryonic density range of interest. In
particular the three models with smaller proton fraction are GM3, Dalen
and NL3, the three models that have the smaller symmetry energy for
$0.05< \rho< 0.1$ fm$^{-3}$.

\begin{figure}
\begin{center}
\includegraphics[scale=0.70]{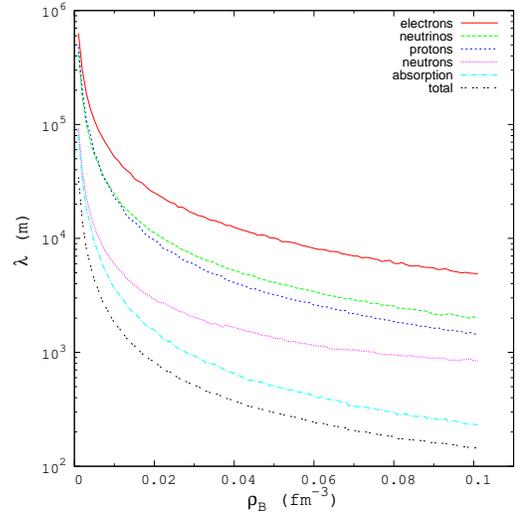}
\end{center}
\caption{Mean-free-path as a function of the total baryonic density using the FSUGold parametrization for 
 the pasta phase matter. Individual contributions for the electrons
 (red), neutrinos (green), protons (blue) and neutrons (purple).
 The absorption contribution is shown in light blue and the total value is shown in black.
 The temperature was taken as $T=1$ MeV and $E_{\nu}=\mu_{\nu}$.}
\label{MeanFST1}
\end{figure}

The individual mean-free-path contribution for each particle type is
shown in Fig.  \ref{MeanFST1} as a function of $\rho_B$,
for the FSUGold and $T=1$Mev.
The curves labeled proton, neutron and neutrino are the contributions for elastic neutral current scattering,
while the curve labeled electron has also an elastic contribution from charged current scattering.
It is clear the dominance of the absorption process for the total cross section.

\begin{figure}
\begin{center}
\includegraphics[scale=0.70]{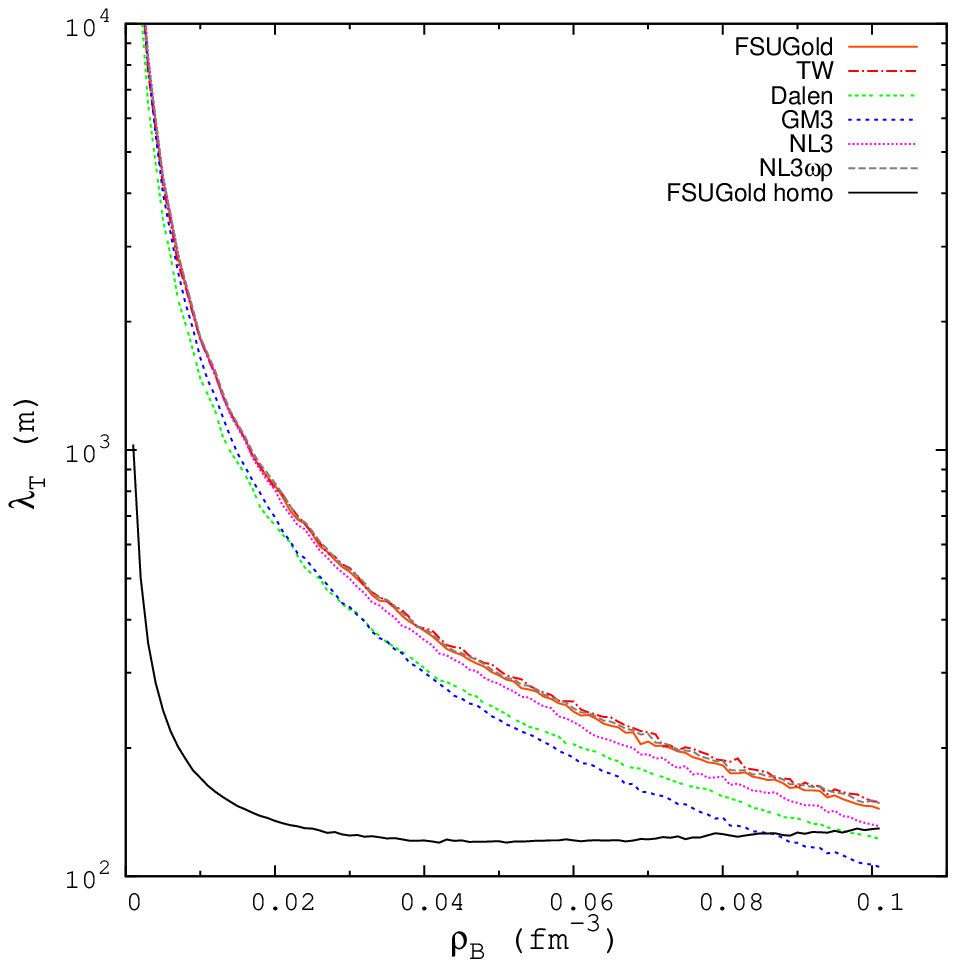}
\end{center}
\caption{Mean-free-path as a function of the total baryonic density comparing various parametrizations for 
 the pasta phase matter. TW (red), Dalen (green), GM3 (blue), NL3
 (pink), NL$\omega\rho$ (grey) and FSUGold (orange). The FSUGold
 for homogeneous matter (black), is also shown. 
 The temperature was taken as $T=1$ MeV and
 $E_{\nu}=\mu_{\nu}$.}
\label{Mean2FST1}
\end{figure}
The sensitivity  of the total mean-free-path to the parametrization 
used is 
presented in Fig.  \ref{Mean2FST1} together with the uniform matter
result obtained with the FSUGold parameter set.
At low densities the models do not differ much. This is expected since all models have similar behaviors,
in particular, the proton fraction is almost the same in all of them,
see \cite{grill12}. The differences occur precisely after
the onset of the non-spherical pasta structures, $\rho_B>0.02~\mbox{fm}^{-3}$, which is a range of
densities that is sensitive to the density dependence of the symmetry
energy. GM3 and Dalen have the lowest mean-free paths:
a smaller proton fraction (see Fig.  \ref{fracFST1}) corresponds to a smaller electron fraction,
which increases the neutrino fraction and the neutrino chemical potential. All these facts
favor the absortion process, and consequently the NMFP is smaller.
Comparing NL3 and
NL3$\omega\rho$ we also conclude that the softer density dependence of the symmetry energy
gives rise to larger mean-free paths.

Considering the scattering of the neutrinos from the individual nucleons of the pasta, as done in the present
calculation, we see that the mean-free path increases a lot
comparatively to the homogeneous matter case, mainly at low densities, when the clusterized matter presents much larger proton
fractions, and, therefore, the absortion process is less likely to occur.
Other reason is that the processes are strongly suppressed by Pauli blocking effects inside the cluster, where
the Fermi energy is larger, while outside the cluster
the space is almost empty. As expected, the larger the neutron and proton densities of the background gas the closer get the pasta
and homogeneous matter mean-free paths.

\begin{figure}
\begin{center}
\includegraphics[scale=0.70]{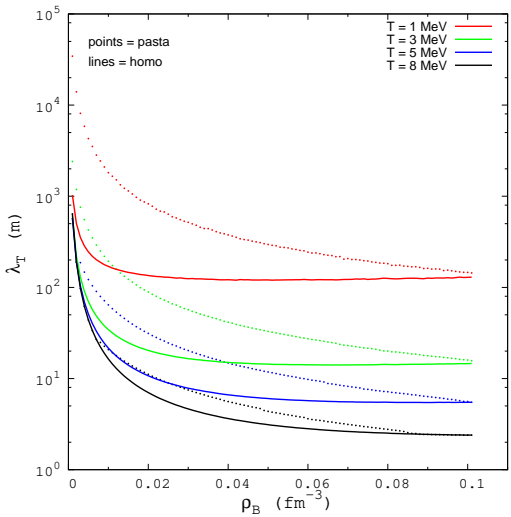}
\end{center}
\caption{The neutrino mean-free-path as a function of the total baryonic density within FSUGold parametrization for 
 the pasta phase (points) compared to homogeneous matter (lines) for
 $T=1$ MeV(red), $T=3$ MeV(green), $T=5$ MeV(blue) and $T=8$ MeV(black)
 and $E_{\nu}=\mu_{\nu}$.}
\label{MeanT135}
\end{figure}

The dependence of the results with the temperature can be seen in Fig. \ref{MeanT135}.
Temperature has a strong effect on the pasta structures which start to melt. Although within a TF
calculation pasta structures still exist at $T>10~MeV$, according to \cite{pethick98}, if thermal fluctuations are
considered the Wigner-Seitz cell structure is supposed to melt for $T>7~MeV$. Temperature increases drastically
the background gas of dripped particles in the Wigner-Seitz cells and, therefore, the larger the temperature the
closer come the pasta mean-free path to the homogeneous matter ones.
The reduction of the mean-free path with the increase of the temperature is also expected, because the
 Fermi-Dirac distributions are smoothed when the temperature increases, making possible more reactions.

\begin{figure}
\begin{center}
\includegraphics[scale=0.70]{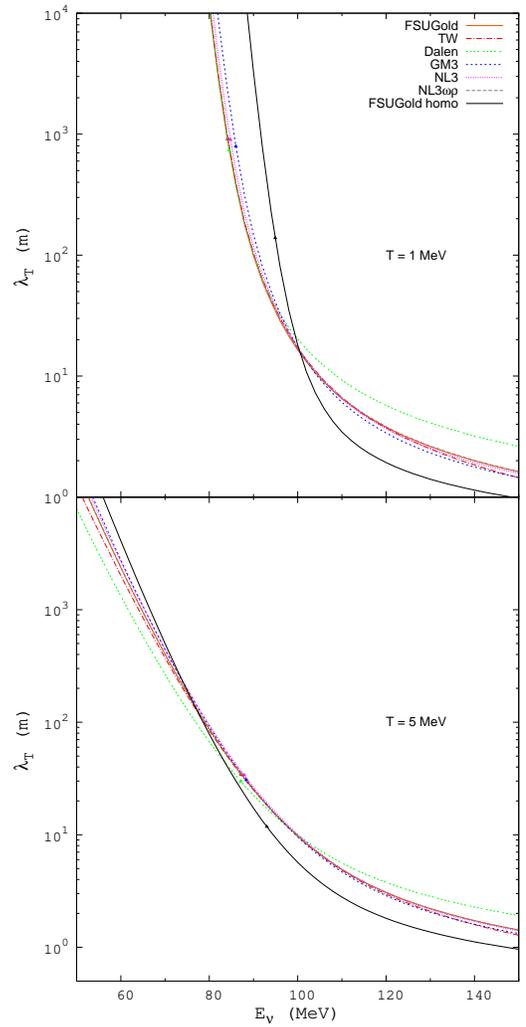}
\end{center}
\caption{Neutrino mean-free-path as a function of the neutrino incident energy for the pasta phase with $\rho_{B}=0.018 \mbox{fm}^{-1}$,
comparing the parametrizations TW(red), Dalen(green), GM3(blue), NL3(pink), NL$\omega\rho$(grey), FSUGold(orange) and FSUGold homogeneous matter(black).
The temperature was taken as $T=1$ MeV (upper panel) and $T=5$ MeV
(lower panel). The triangles indicate the NMFP at the neutrino
chemical potential, $\mu_\nu$, of the respective model.}
\label{MeanET1}
\end{figure}
\begin{figure}
\begin{center}
\includegraphics[scale=0.70]{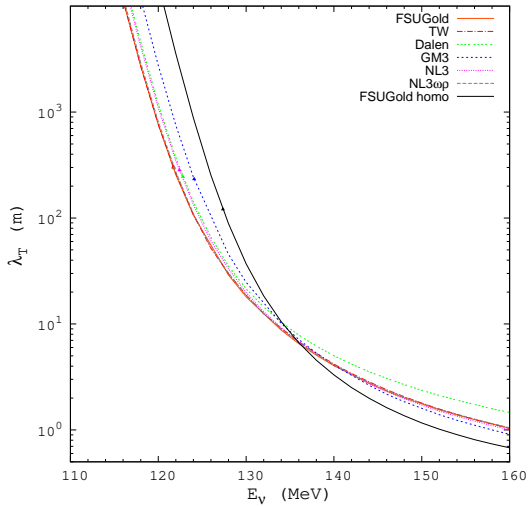}
\end{center}
\caption{Neutrino mean-free-path as a function of the neutrino
  incident energy for the pasta phase with $\rho_{B}=0.05 \mbox{fm}^{-1}$ and
  $T=1$ MeV, 
comparing the parametrizations TW(red), Dalen(green), GM3(blue), NL3(pink), NL$\omega\rho$(grey), FSUGold(orange) and 
FSUGold homogeneous matter(black). The triangles indicate the NMFP at the neutrino
chemical potential, $\mu_\nu$, of the respective model.}
\label{MeanET1-2}
\end{figure}

In Fig. \ref{MeanET1} the NMFP is plotted as a
function of the neutrino incident energy for the pasta phase with
$\rho_{B}=0.018 \mbox{fm}^{-1}$. We also include the homogeneous matter
result obtained within FSUGold. With triangles, we select the values
of $\lambda$ at the neutrino chemical potential,
at the two temperatures shown, $T=1$ and $5$ MeV. We expect that the neutrinos
with energy around the chemical potential will provide the main contribution to the dispersion of energy in the system,
because at low temperatures the system is practically degenerate and Pauli blocking factors will weaken other contributions.
The highly degenerate regime is
expected for $\mu_i/T \gg 1$, and this condition is essentially true
for temperatures below the pasta melting temperature. The differences
between the two temperatures is mainly explained by the drip of nucleons out of the
clusters increasing the background gas, and reducing the differences
between the pasta and homogeneous matter results. Temperature also
decreases the mean-free path by almost one order of magnitude both for
the pasta and the homogeneous matter calculation and this is explained
by the opening of new  transitions that are Pauli blocked at $T$ close to zero.

In Fig. \ref{MeanET1-2} we consider a  larger density corresponding to
the slab geometry at $T=1$ MeV. For these densities the gas of dripped
neutrons is denser and the proton fraction smaller (see
 the blue region in Fig. \ref{fracFST1} in comparison with
 the red region). Consequently, the NMFP decreases and
 comes closer to the homogeneous matter result.

\section{Conclusions}\label{conclusions}

We have studied the effect of the pasta phase,  occurring in the inner
crust of a neutron star, on  the neutrino mean-free-path (NMFP). 
The pasta phases have been obtained within a
self-consistent Thomas-Fermi approximation, as explained in references
\cite{tomas,sidney}.
Several relativistic nuclear models have been used
to describe the pasta phase, both  with nonlinear meson terms and constant couplings
, and with density-dependent coupling constants. In
particular, we were interested in discussing whether the properties of
the models, such as the density dependence of the symmetry energy, would have some influence on the NMFP. 
We have also studied the effect of the temperature. It should be
stressed, however, that the present work is restricted to the density
and the temperature range for which the pasta phases exist,
e.g. $T\lesssim 8$ MeV and $\sim 0.0002 \mbox{fm}^{-3}<\rho< \rho_t$ where the density at
the crust-core transition $\rho_t\sim \rho_0/2.$ 
We have considered both charged and neutral current reactions. It has been shown that the absortion process dominates the NMFP, but other
processes can not be discarded.

At low density, $\rho_B<0.02 \mbox{fm}^{-3}$, where the pasta phases obtained within
the different parametrizations have similar
properties \cite{grill12}, the NMFP has a small dependence on  the parametrization. As the
density increases, namely at the layers close to the upper boarder of
the inner crust, the dependence on the parametrization becomes more
important.
With a larger proton fraction, the absortion process is less likely to occur.
Consequently, models with larger proton
fraction result in larger NMFP, and since the proton fraction is
closely related to the density dependence of the symmetry energy, its
effect on the NMFP  is not negligible.
We have shown that the NMFP is larger in the presence of
pasta phases than considering homogeneous $\beta$-equilibrium matter
at the same density.
Besides the fact that the pasta phases have a larger proton fraction,
the processes are strongly suppressed by Pauli blocking effects inside the cluster, where
the Fermi energy is larger, while outside the cluster
the space is almost empty. When the density increases, the closer get the pasta
and homogeneous matter mean-free paths.

As the temperature increases, the pasta phase starts to melt and the
pasta phase EOS approaches the
infinity nuclear matter EOS, and, therefore,  the NMFP in both cases get closer.
At the temperatures we have considered the NMFP neutrinos are
degenerate, and, therefore, the Pauli blocking factors ensure that only
neutrinos close to the Fermi energy are involved
at the dispersion of energy in the system.
Comparing with the
homogeneous case, we expect that  neutrinos with less energy are
involved in the presence of pasta phases, which give rise to larger NMFP.

The NMFP as function of the neutrino energy for the different models
do not differ much at low densities and temperatures, because the models have
similar properties in this region of phase space. With the increase of the temperature and density, the
dependence on the models starts to appear. The great variation of the NMFP with
the neutrino energy around the neutrino chemical potential is
attributed to 
Pauli blocking effects.

Our results imply that the effects of the pasta phase can not be neglected when
calculating the NMFP at baryonic densities below $0.1 \mbox{fm}^{-3}$ and temperatures
below $10 MeV$. The parametrization used to described the pasta phase  has also
influence on the results, in particular due to the density dependence of the
symmetry energy.

We have also not considered coherent scattering of the neutrinos
from the pasta clusters as done in \cite{horowitz2005,sonoda2007,prakash99}.  These processes are
important when all neutrons respond coherently, for scattering by
neutrinos with a wavelength of the size of the cluster, or energies
$E_\nu \lesssim 75$ MeV. The scattering processes discussed in the
present work become important when the inner constitution of the clusters
is distinguished by the neutrinos, and, therefore, for neutrino with an energy above that value.

\section*{ACKNOWLEDGMENTS}

We acknowledge partial support from CAPES and CNPq. This work is partly supported by the project PEst-OE/FIS/UI0405/2014 developed under the inititative QREN financed by the UE/FEDER through the program
COMPETE-``Programa Operacional Factores de Competitividade''.

\section{Appendix}\label{appendix}

 According to equation (\ref{secao nucleon}), with $\boldsymbol{k}_1=k_1 \hat{z}$ and after
the $\boldsymbol{k}_4$ integration in the scattering case:

\begin{align} \label{secao nucleon 2}
\sigma_{scat} =& \f{{G_F}^4}{(2\pi)^5}  \int d^3 r \int {d^3 k_2} \int_0^\pi sen(\theta_3){d \theta_3} \int_0^{2\pi} {d \phi_3}\,
k_3^2 \cdot\nonumber\\
&\f{\eta_2(T) (1-\eta_3 (T)) (1-\eta_4 (T))}{|\boldsymbol{v}_1 - \boldsymbol{v}_2|} \cdot\nonumber\\
& \f{k_1 + E^*_2 -k_3}{k_1(1-cos(\theta_3))+E_2^{*}-k_2cos(\theta_{23})} \cdot\nonumber\\
& \{(c_V+c_A)^2(1-v_2cos(\theta_{2}))(1-f_{123}) \nonumber\\
& +(c_V-c_A)^2(1-v_2cos(\theta_{23})(1-g_{123}) \nonumber\\
& - \f{M^*_2 M^*_4}{E_2^* E_4^*}(c_V^2-c_A^2)(1-cos(\theta_{3})) \}
\end{align}
with
$$
f_{123}=\f{k_1cos(\theta_3)+k_2cos(\theta_{23})-k_3}{E^*_4};
$$
$$
g_{123}=\f{k_1+k_2cos(\theta_2)-k_3cos(\theta_3)}{E^*_4};
$$
$$
k_3 = \f{k_1\left(E^*_2-k_2 cos(\theta_2)\right)}{k_1 + E^*_2 - k_1 cos(\theta_3) - k_2 cos(\theta_{23})};
$$

For the absorption, we find:
\begin{align} \label{secao neutron 2}
\sigma_{abs} =& \f{{G_F}^4}{(2\pi)^5}   \int d^3 r \int {d^3 k_2} \int_0^\pi sen(\theta_3){d \theta_3} \int_0^{2\pi} {d \phi_3} k_3^2\cdot\nonumber\\
&\f{\eta_2(T) (1-\eta_3 (T)) (1-\eta_4 (T))}{|\boldsymbol{v}_1 - \boldsymbol{v}_2|} \cdot\nonumber\\
& \f{\sqrt{m_e^2+k_3^2}(C-\sqrt{m_e^2+k_3^2})}{C k_3 - F \sqrt{m_e^2+k_3^2}} \cdot\nonumber\\
& \{(g_V+g_A)^2(1-v_2cos(\theta_{2}))(1-f_{123}) \nonumber\\
& +(g_V-g_A)^2(1-g_{123})(1-v_2cos(\theta_{23})) \nonumber\\
& - \f{M^*_2 M^*_4}{E_2^* E_4^*}(g_V^2-g_A^2)(1-cos(\theta_{3}) \}
\end{align}
where now
\begin{align}
k_3 =& -\f{2F(C^2+m_e^2-D)}{4(k_1 cos(\theta_3)+k_2 cos(\theta_{23}))^2-4C^2}\nonumber\\
&- \f{\sqrt{16m_e^2C^2F^2+4C^2(C^2+m_e^2-D)^2-16m_e^2C^4}}{4F^2-4C^2};\nonumber
\end{align}
\begin{align}
&D = (\boldsymbol{k}_1 + \boldsymbol{k}_2)^2 + M^*_4;\nonumber\\
&C = k_1 + E^*_2 - g_\rho b_0;\nonumber\\
&F = k_{1}cos(\theta_3)+k_{2}cos(\theta_{23}).
\end{align}

The constants are:

\vspace{0.5cm}

neutrino-proton: $c_V=1/2-2sen^{2}(\theta_w)$; $c_A=1.23/2$

neutrino-neutron: $c_V=-1/2$; $c_A=-1,23/2$

neutrino-electron: $c_V=1/2+2sen^{2}(\theta_w)$; $c_A=1/2$

neutrino-neutrino: $c_V=\sqrt{2}$; $c_A=\sqrt{2}$

absorption : $g_V=C$; $g_A=-1.23C$

\vspace{0.5cm}

\noindent with $C = 0.973$ and $sen^{2}(\theta_w)= 0.230 $. The collisions
neutrino-electron and neutrino-neutrino can be represented by two (first-order) Feynman diagrams each one, in such a way that the values
of $c_V$ and $c_A$ can be re-defined to accommodate the different cross section contributions in a single expression, as given above. Also, we define:

\beq
E^* = \sqrt{\boldsymbol{k}^2 + M^{* 2}}. \nonumber
\eeq

For the particle energy  we have for the nucleon:

$$
E_i = E^* + \Gamma_v V_0(\boldsymbol{r}) \pm \f{1}{2}\Gamma_{\rho} b_0(\boldsymbol{r}) + e A_0(\boldsymbol{r}),
$$
in the NL parametrization case and
$$
E_i = E^* + \Gamma_v V_0(\boldsymbol{r}) \pm \f{1}{2}\Gamma_{\rho} b_0(\boldsymbol{r}) + e A_0(\boldsymbol{r})
$$
\beq + \f{\partial \Gamma_v}{\partial \rho_B} \rho_B V_0
+ \f{\partial \Gamma_\rho}{\partial \rho_B} \f{\rho_3}{2} b_0 + \f{\partial \Gamma_\delta}{\partial \rho_B} \rho_{s3} \delta_0
- \f{\partial \Gamma_s}{\partial \rho_B} \rho_s \phi_0, \nonumber
\eeq

\noindent in the DD case. The plus sign in the equations above has to be chosen for the proton and
the minus sign for the neutron. Also, $V_0$, $\phi_0$, $b_0$ and $\delta_0$ are the time-like components of the meson fields. For the electron:

\beq
E_e = \sqrt{k^2 + m_e^2} - e A_0(\boldsymbol{r}); ~ M^{*}\rightarrow m_e \nonumber
\eeq

\noindent and for the neutrino:
\beq
E_\nu = k ; ~ M^{*}\rightarrow 0. \nonumber
\eeq

Finally, in order to take in to account the finite size of the nucleon, we have multiplied the constants $g_V$ ($g_A$), or
$c_V$ ($c_A$), by the corresponding form factor.  Using the parametrizations as explained in \cite{INPC2013}, we have taken
a common structure factor given by:

\beq
\left(1+\frac{4.97q^2}{4M^2} \right)^{-2},
\eeq 

\noindent where $q\equiv|\boldsymbol{q}|=|\boldsymbol{p_1}-\boldsymbol{p_3}|$.

\end{document}